\documentclass[useAMS,usenatbib]{mn2e}

\title[Nulls subpulse drift and mode-switching]{Nulls subpulse drift and
mode-switching in pulsars: the polar-cap surface}
\author[P. B. Jones]{P. B. Jones\thanks{E-mail:p.jones1@physics.ox.ac.uk} \\
Department of Physics, University of Oxford, Denys Wilkinson Building,\\
Keble Road, Oxford OX1 3RH, England}
\begin{document}

\date{}

\pagerange{\pageref{firstpage}--\pageref{lastpage}} \pubyear{}

\maketitle

\label{firstpage}

\begin{abstract}
Little attention has so far been paid to the division of the observed population %%@
between pulsars of the two spin directions that are possible.  Almost all pulsars with %%@
positive corotational charge density at the polar caps are expected to satisfy %%@
space-charge limited flow boundary conditions.  Charge separation by blackbody %%@
photo-electric transitions in moving ions limits the acceleration potential, %%@
analogously with the more usually considered pair creation.  But the limitation is more %%@
severe so that proton and ion energies can be relativistic but not ultra-relativistic, %%@
and these allow the growth of Langmuir-mode induced turbulence that couples directly %%@
with the radiation field, as shown by Asseo, Pelletier \& Sol.  The consequences of %%@
this, and of the several possible physical states of the polar cap, are described, %%@
qualitatively, as possible explanations for the complex phenomena of nulls, subpulse %%@
drift and mode-switching observed in sub-sets of pulsars.
\end{abstract}


\section{Introduction}

The past twenty years have seen a number of developments that are likely to be relevant %%@
to pulsar radio and X-ray emission.  Apart from the many improvements in the quality %%@
and scope of observations, there has been the recognition of the role of the %%@
Lense-Thirring effect in acceleration at the polar cap (Muslimov \& Tsygan 1992) and  %%@
of the significance of plasma turbulence (see the review of Melrose 2000). Most of the %%@
very many theoretical papers published have assumed a relative orientation of rotation %%@
spin ${\bf \Omega}$ and polar-cap magnetic flux density ${\bf B}$ such that ${\bf %%@
\Omega}\cdot{{\bf B}} > 0$ with Goldreich-Julian charge density $\rho_{GJ} < 0$ and %%@
electron acceleration.  In this case, there can be no doubt that the electron work %%@
function is so small that the space-charge limited flow boundary condition on the %%@
corotating-frame electric field, ${\bf E}_{\parallel} = 0$, is satisfied on the %%@
polar-cap surface at all instants of time. (The subscripts $\parallel$ and $\perp$ %%@
refer to directions locally parallel with and perpendicular to ${\bf B}$.)  There %%@
appears to be no reason why neutron stars with ${\bf \Omega}\cdot{\bf B} < 0$ and %%@
$\rho_{GJ} > 0$ should not be present in the pulsar population, but in this case, the %%@
best existing calculations of ion separation energies (see Medin \& Lai 2006) combined %%@
with estimates of the polar-cap temperatures can be used to obtain values of the %%@
minimum magnetic field for which the boundary condition ${\bf E}_{\parallel} \neq 0$ is %%@
possible. These are typically of the order of $10^{14}$ G (Jones 2011) and are indirect %%@
evidence that all except possibly a very small number of pulsars with spin direction %%@
such that ${\bf \Omega}\cdot{\bf B} < 0$ satisfy the space-charge limited flow boundary %%@
condition rather than the ${\bf E}_{\parallel} \neq 0$ condition assumed in the classic %%@
polar-cap model of Ruderman \& Sutherland (1975).

The concept of localized regions of electron-positron pair production, referred to as %%@
sparks, moving over the neutron-star surface within the polar cap in an organized way, %%@
has been adapted from the Ruderman-Sutherland model and the ${\bf E}_{\parallel} \neq %%@
0$ boundary condition, but is now widely used phenomenologically, with no reference to %%@
boundary conditions or to the sign of ${\bf \Omega}\cdot{\bf B}$, in describing %%@
observations such as nulls and subpulse drift.
In the ${\bf \Omega}\cdot{\bf B} > 0$ case, with space-charge limited flow, no physical %%@
basis for such a phenomenological description has so far been published, the reason %%@
being that electrons are the only possible negatively-charged particles for outward %%@
acceleration and the space-charge limited flow boundary condition ${\bf E}_{\parallel} %%@
= 0$ can be satisfied at any conceivable polar-cap surface temperature.  The state of %%@
the accelerated plasma is then simply a problem in electromagnetism with defined %%@
boundary conditions.
However, for ${\bf \Omega}\cdot{\bf B} < 0$, it has been shown that subpulse formation %%@
and movement, either organized or disorganized, over the polar-cap surface is quite %%@
natural and should be observable in most radio pulsars with this spin direction (Jones %%@
2010a, 2011; hereafter Papers I and II).  The basis for this conclusion is an %%@
examination of the physical processes occurring in electromagnetic showers formed at %%@
the polar-cap surface by photo-electrons accelerated inward.  The problem of finding %%@
the state of the accelerated plasma is not restricted to electromagnetism but also %%@
depends on proton production by the formation and decay of the giant dipole resonance %%@
in the electromagnetic showers.

Subpulse modulation (the wide variations of intensity at a fixed longitude in a %%@
sequence of observed pulses) has been recognized as an almost universal property of %%@
pulsars ever since their discovery. Weltevrede, Edwards \& Stappers (2006) found it %%@
present in 170 out of a sample of 187 pulsars selected only by signal-to-noise ratio, %%@
evidence that its presence is very likely to be independent of the sign of ${\bf %%@
\Omega}\cdot{\bf B}$. Strong plasma turbulence is now viewed as the most probable %%@
source of radio emission (see Melrose 2000).  Its development from a quasi-longitudinal %%@
Langmuir mode and the possible formation of a random array of stable Langmuir solitons %%@
has been investigated by Asseo, Pelletier \& Sol 1990, and by Asseo \& Porzio 2006. It %%@
has been shown recently (Jones 2012; hereafter Paper III) that the formation of this %%@
mode is also possible in a two-component beam consisting of protons and relativistic %%@
but not ultra-relativistic ions that can be formed as a consequence of the reverse flow %%@
of photo-electrons to the polar cap.  Thus it is not surprising that subpulse %%@
modulation is almost universal.
That a secondary low-energy electron-positron plasma is not the only possible source of %%@
plasma turbulence is also consistent with the fact that common characteristics of %%@
coherent radio emission are observed in the pulsar population even though the inferred %%@
polar-cap magnetic flux densities vary by five orders of magnitude.  Pulsars with %%@
either sign of ${\bf \Omega}\cdot{\bf B}$ are therefore expected to produce individual %%@
pulses of similar form although if there is no pair production,
there may be differences that are difficult to detect such as in the distributions of %%@
spectral index caused by the different plasma frequencies and Lorentz factors in the %%@
two cases.  But the present paper (see also Paper II) proposes that   phenomena such as %%@
nulls, mode-changes and organized sub-pulse motion are restricted to the ${\bf %%@
\Omega}\cdot{\bf B} < 0$ case.  These involve time-scales that have no obvious %%@
explanation in the ${\bf \Omega}\cdot{\bf B} > 0$ direction of spin.

Existing observations show that subpulse drift is a frequent though not universal %%@
phenomenon. Weltevrede, Edwards \& Stappers found that 68 of their sample of 187 %%@
pulsars show observable subpulse drift.  There appears to be no published survey either %%@
of nulls or of mode-changes based on a carefully defined sample as in the above work.  %%@
But the most recent and largest tabulation of data (Wang, Manchester \& Johnston 2007) %%@
shows that nulls must be a moderately frequent phenomenon.  Systematic information %%@
about mode-changing is much more sparse (see Kramer et al 2006).

Models of subpulse formation and drift recently developed (see, for example, Gil \& %%@
Sendyk 2000) are based on the classic paper of Ruderman \& Sutherland (1975) and assume %%@
the ${\bf \Omega}\cdot{\bf B} < 0$ case with the polar-cap surface boundary condition %%@
${\bf E}_{\parallel}\neq 0$.  These have all the attributes of a good physical model, %%@
particularly simplicity and predictive capability in that it is possible to write down %%@
an expression for the ${\bf E}\times{\bf B}$ subpulse drift velocity.  
But the boundary condition requires surface magnetic flux densities approximately two %%@
orders of magnitude larger than the dipole fields inferred from the spin-down rates.  %%@
Although the presence of highly complex field geometries formed from higher-multipole %%@
components cannot be excluded, it would be strange if they were present with the %%@
frequency indicated by the survey of Weltevrede, Edwards \& Stappers.  Consequently, %%@
Papers I and II investigated physical processes at the polar-cap surface and have shown %%@
that the composition of the plasma accelerated under the space-charge limited flow %%@
(SCLF) boundary condition is usually neither time-independent nor uniform over its %%@
whole area.  Localized areas in which the accelerated plasma is suitable for the growth %%@
of the quasi-longitudinal Langmuir mode exist naturally and can move in either an %%@
organized or disorganized way, so acting as sources for subpulses with the same %%@
properties.  Thus plasma systems analogous with Ruderman-Sutherland sparks are present %%@
under the ${\bf E}_{\parallel} = 0$ boundary condition. But we emphasize that their %%@
motion is unrelated with ${\bf E}\times{\bf B}$ drift but is determined by the time %%@
$\tau _{p}$ in which electromagnetic shower protons diffuse to the surface.

This paper attempts to present a description of the early stages of the coherent %%@
emission mechanism that is very different from the canonical.  In particular, the %%@
unexpected conclusion of Paper III that relatively low-energy ion and proton beams can %%@
be formed means that it is necessary to re-assess the model of sub-pulse drift %%@
discussed in Paper II.  Polar caps of pulsars with spin ${\bf \Omega}\cdot{\bf B} < 0$ %%@
are seen to be very complex systems compared with those of the electron acceleration %%@
${\bf \Omega}\cdot{\bf B} > 0$ case.  But we must presume that both are present in the %%@
neutron-star population and so it is worth examining how the properties of the observed %%@
pulsar population may be related with spin direction.

Before proceeding further, it must be admitted that the work has at least one drawback.  %%@
Processes of the kind considered in Papers I and II, occurring at a continuously %%@
changing real condensed-matter surface, are not necessarily susceptible to the %%@
formulation of simple physical theories.  The behaviour of such systems can be varied %%@
and complex.  But this is not inconsistent with the observed properties of many %%@
individual pulsars.

Section 2 summarizes the assumptions about the properties of the open region of the %%@
magnetosphere that we are obliged to make in both qualitative and quantitative work.
Strong plasma turbulence is possible only for limited intervals of particle energy at %%@
altitudes $z$ above the polar-cap surface smaller than $z \sim 10R$, where $R$ is the %%@
neutron-star radius. This is described in more detail in Section 3. Section 4 draws on %%@
the work of Paper III to obtain values of the parameter $K$, which is the number of %%@
protons produced by photo-electrons per unit nuclear charge accelerated. We then give
in Section 5 a description of the polar cap showing how subpulses are formed and that %%@
either chaotic or organized motion is possible.  An attempt is made to show that this %%@
model of the polar cap provides a physical basis for phenomena such as nulls, subpulse %%@
drift, and mode changes.

\section{The open magnetosphere at the polar cap}

At the base of the magnetosphere there is an atmosphere of ions, assumed to be in local %%@
thermodynamic equilibrium (LTE), having a scale height of the order of $10^{-1}$ cm.
Its total mass is temperature-dependent but is broadly equivalent to $10^{-1} - 10^{1}$ %%@
radiation lengths.  Thus it contains only the earlier part of the electromagnetic %%@
showers formed by the
reverse-electron flux.  There is fractionation of the ionic charge to mass ratio. Ions %%@
with the highest value are concentrated at the top of the atmosphere and so %%@
preferentially enter the acceleration region.  The proton number density is many orders %%@
of magnitude smaller than that of the ions and consequently
has no effect on the structure of the atmosphere.  Protons are not in static %%@
equilibrium within the LTE ion atmosphere but are subject to an outward-directed net %%@
gravitational and electrostatic force. Their motion is therefore a combination of %%@
diffusion and drift with a characteristic time $\tau _{p}$ elapsing between formation %%@
and arrival at the top of the atmosphere. If the rate of formation is such that the %%@
Goldreich-Julian flux would be exceeded, the excess protons form a thin atmosphere in %%@
equilibrium at the top of the LTE ion atmosphere.  We refer to Sections 2, 3 and 5 of %%@
Paper II for a more complete account of these topics. Values $\tau _{p}\sim 10^{-1} - %%@
10^{0}$ s are estimated.

The polar-cap radius adopted here and in Paper III is based on dipole-field geometry %%@
and is that given by Harding \& Muslimov (2001),
\begin{eqnarray}
u_{0}(0) = \left(\frac{2\pi R^{3}}{cPf(1)}\right)^{1/2},
\end{eqnarray}
for rotation period $P$,
in which $f(1) = 1.368$ for the neutron star of mass $1.4M_{\odot}$ and radius $R = 1.2 %%@
\times 10^{6}$ cm.  As in Paper III, we use the fact that the open magnetic flux lines %%@
are contained within a narrow tube, assumed to be of circular cross-section
whose radius, $u_{0}(z) = u_{0}(0)\eta^{3/2}$ for a dipole field, increases only very %%@
slowly as a function of altitude on the magnetic axis $z = (\eta - 1)R$ at radii $\eta %%@
\sim 10$ or smaller. Therefore, we can adequately represent a small section of the tube %%@
at altitude $z$ as a right cylinder of locally-defined radius
${\bf u} = {\bf u}_{0}(z)$.  Then the electrostatic potential $\Phi$ at $z\gg u_{0}(0)$ %%@
and under the SCLF boundary conditions is approximately
\begin{eqnarray}
\Phi(u,z) = \pi\left(u_{0}^{2}(z) - u^{2}\right)\left(\rho(z) - \rho _{GJ}(z)\right),
\end{eqnarray}
in terms of the difference between the charge density $\rho$ and the Goldreich-Julian %%@
charge density $\rho _{GJ}$ which is almost constant over the cross-sectional area of %%@
the tube.  We refer again to Harding \& Muslimov (2001) for these quantities at radial %%@
coordinate $\eta$,
\begin{eqnarray}
\rho _{GJ} = -\frac{Bf(\eta)}{cP\alpha \eta^{3}f(1)}\left(1 - \frac{\kappa}{\eta^{3}}
\right)\cos\psi
\end{eqnarray}
and,
\begin{eqnarray}
\rho = - \frac{Bf(\eta)}{cP\alpha \eta^{3}f(1)}\left(1 - \kappa \right)\cos\psi,
\end{eqnarray}
in which $B$ is the surface magnetic flux density.  Equation (4) gives the SCLF charge %%@
density undisturbed by photo-electric transitions or electron-positron pair creation.
In these expressions, $\kappa$ is the dimensionless Lense-Thirring factor which we %%@
assume to be $\kappa = 0.15$, and $\psi$ is the angle between ${\bf \Omega}$ and ${\bf %%@
B}$.  Equation (4) also assumes that particle velocities, including those of protons %%@
and ions, differ negligibly from the velocity of light.
The function $f(\eta)/\alpha f(1)$ including the red-shift factor $\alpha$ is a %%@
slowly-varying function of $\eta$ and is approximately unity at low altitudes.
Thus $\rho _{GJ}(0) = (1 - \kappa)\rho _{GJ}(\infty)$ where $\rho _{GJ}(\infty)$ is the %%@
flat space-time Goldreich-Julian charge density. 
Equations (3) and (4) exclude terms that can become significant at higher altitudes for %%@
which our geometrical model also fails, but nevertheless equation (2) is adequate for %%@
the  estimation of proton and ion energies and of photo-electric transition rates. 
It is convenient to use the energy units of particle physics for polar-cap processes %%@
and therefore we define a potential energy difference $V({\bf u},z) = -e\Phi$.
It has the value $V_{max}({\bf u},z)$ on a flux line with polar-cap coordinate ${\bf %%@
u}$ for the undisturbed SCLF charge density of equation (4). It is also useful to have %%@
a simple expression for $V_{max}$ on the magnetic axis,
\begin{eqnarray}
V_{max}(0,\infty) \approx  \frac{2\pi^{2}R^{3}\kappa eB}{c^{2}f(1)P^{2}}
  = 1.25\times 10^{3}\frac{B_{12}}{P^{2}} \hspace{5mm}{\rm GeV},
\end{eqnarray}
for acceleration above the polar cap, where $B_{12}$ is the surface magnetic flux %%@
density in units of $10^{12}$ G.

The above statement is subject to a serious qualification concerning the definition of %%@
the open magnetosphere.  If many secondary low-energy electron-positron pairs were %%@
created per primary particle accelerated, there would be no difficulty in assuming the %%@
unconstrained outward flow of primary particles and secondary pairs on any magnetic %%@
flux line intersecting the light cylinder.  The separate velocity distributions of %%@
secondary electrons and positrons would have the capacity to adjust so as to maintain %%@
the condition ${\bf E}_{\parallel} = 0$ at all points in the open region.  But there %%@
must be serious doubts about the universal existence, in the pulsar population, of the %%@
necessary density of secondary-pair plasma. Apart from ${\bf \Omega}\cdot{\bf B} < 0$ %%@
pulsars in which there may be no pair creation, calculations of the secondary pair %%@
density produced by the inverse Compton scattering (ICS) of polar-cap photons by %%@
outward accelerated electrons in the ${\bf \Omega}\cdot{{\bf B}} > 0$ case (Hibschman %%@
\& Arons 2001; Harding \& Muslimov 2002) predict in many pulsars pair densities that %%@
are small compared with unity, per primary electron.  It is true that more complex %%@
field geometries, such as the off-set dipole, are possible and may enhance pair %%@
production (see Harding \& Muslimov 2011) but there remain at least some hints of a %%@
problem in the pair density.  In the absence of a pair plasma with the required %%@
properties, motion 
in the magnetosphere beyond the $\eta\sim 10$ region depends on the sense of curvature %%@
of the open flux line: there is further acceleration for $|\cos\psi|$ increasing, but %%@
deceleration in the opposite case which is capable of cancelling the acceleration at %%@
lower altitudes.  A section of the polar cap is then a dead zone from which there is no %%@
outward particle flux.  Whilst the change in active polar-cap shape is not very %%@
significant, the potential given by equation (2) is reduced, very roughly, in %%@
proportion to the decrease in area.  The component ${\bf E}_{\parallel}$ in the dead %%@
volume is very small and is that required to maintain a charge-separated equilibrium %%@
with gravity and the centrifugal force.

A further source of uncertainty is the consistency with which the total electric charge %%@
of the neutron star remains constant. The rate of loss of charge at the polar caps is
$\dot{q} = 2\pi u^{2}_{0}(0)\rho _{GJ}(0)c$ whose compensation is likely to be easy if %%@
there is strong secondary pair creation, but is otherwise obscure.  An order of %%@
magnitude estimate of the relevant time-scale can be found by dividing a typical %%@
light-cylinder field strength by the rate of change of the light-cylinder radial %%@
electric field given by $\dot{q}$.  This is,
\begin{eqnarray}
\left(\frac{BR^{3}}{R^{3}_{LC}}\right)\left(\frac{4\pi R^{2}_{LC}}{\dot{q}}\right)
\sim 2P  \hspace{2mm} {\rm s},
\end{eqnarray}
where $R_{LC}$ is the light-cylinder radius.
This indicates that if there is to be no substantial change in the structure of the %%@
magnetosphere, cessation of the processes maintaining charge constancy is possible only %%@
for time intervals no more than of the order of $P$.

But the great uncertainty affecting ${\bf \Omega}\cdot{\bf B} < 0$ pulsars is in the %%@
atomic number of matter at the neutron-star surface.  If this were small $(Z\sim 6)$ %%@
there would be a negligible reverse-electron flux and, as we shall see in Section 3, no %%@
coherent radio emission unless curvature radiation electron-positron pair production is %%@
possible.  There is some indirect observational evidence for $Z = 6$ (Ho \& Heinke %%@
2009; see also Bogdanov \& Grindlay 2009, Yakovlev et al 2011, and Zhu et al 2009) but %%@
it does not appear strong.  Theoretical work does indicate the possibility of atomic %%@
numbers other than the canonical $Z = 26$.  Calculations by Hoffman \& Heyl (2009) for %%@
isolated neutron stars with no mass accretion show the presence of $Z = 14$, though %%@
with small total mass, but Pearson, Goriely \& Chamel (2011) find a range of values $Z %%@
> 26$.  The mass per unit area removed by a Goldreich-Julian current density of nuclei %%@
with mass number $A$ in an ${\bf \Omega}\cdot{\bf B} < 0$ pulsar in a life-time $t$ is,
\begin{eqnarray}
\frac{Am_{p}}{Z}\int^{t}_{0}\frac{-B\cos\psi}{eP(t^{\prime})}dt^{\prime} \approx %%@
\frac{-2Am_{p}Bt\cos\psi}{ZeP(t)},
\end{eqnarray}
assuming that $B$ and $\psi$ are time-independent.  This is $\sim 4\times %%@
10^{11}B_{12}P^{-1}$ g cm$^{-2}$ at $t = 10^{6}$ yr, and over the whole surface of the %%@
star amounts to a layer of no more than $4\times 10^{-9}B_{12}P^{-1}$ $M_{\odot}$.  It %%@
is a quantity that could easily be the result of unknown details of the formation %%@
process such as fall-back.  In view of this, direct experimental evidence of the nuclei %%@
that are actually present at the polar-cap surface is clearly needed.

The uncertainty in the atomic number of polar-cap surface nuclei has an obvious effect %%@
on photo-electric transitions, and therefore on the reverse-electron flux and the %%@
acceleration potential difference for particles moving on any given magnetic flux line.
The dependence of $\Phi$ on $\rho(z)$ in equation (2) means that the production of %%@
oppositely charged particles in a charge-separated region at altitude $z$ under the %%@
SCLF boundary condition tends to cancel the local electric field ${\bf E}_{\parallel}$.  %%@
This is well-known in the case of electron-positron pairs but is, of course, also true %%@
for photo-electrons (see Paper III).

Equation (2) is not valid at low altitudes $z \sim u_{0}(0)$. Near the polar-cap %%@
surface, $z \ll u_{0}(0)$,
${\bf E}_{\parallel}$ is the inertial acceleration field first described by Michel %%@
(1974).  The problem of finding the electric field is one-dimensional in $z$ and the %%@
inertial ${\bf E}_{\parallel}$ is not large.  Its contribution to particle acceleration  %%@
is small compared with that of the Muslimov-Tsygan effect.  
We refer to Paper III for an approximate model for this region.

The polar-cap and whole-surface neutron-star temperatures are treated as distinct in %%@
Paper III and in the present work.  Calculated photo-electric transition rates for %%@
polar-cap blackbody photons are negligibly small above an altitude
$z = h \approx 4u_{0}(0)$, principally because the photon momentum component %%@
perpendicular to ${\bf B}$ becomes small at higher altitudes.  This is unfavourable %%@
both for the Lorentz transformation to the rest-frame of the accelerated ion and for %%@
the cross-section to the lowest Landau state of the emitted electron (see Paper III).

The SCLF boundary conditions lead to an electric field having a complex form in the $z %%@
< h$ region which is unlikely to be well represented by equation (2).  There also exist %%@
the complications, just discussed, of the inertial ${\bf E}_{\parallel}$ and of %%@
photo-electric transitions in this interval.  But the boundary conditions also limit %%@
the extent to which $\rho(z)$ can change if the condition $E_{\parallel} > 0$ is to be %%@
satisfied.  Thus the difference between the mean ion charges $Z_{h}$ at $z = h$ and %%@
$\tilde{Z}$ in the assumed local thermodynamic equilibrium of the atmosphere at $z = 0$ %%@
cannot be large.  Adjustments in $\rho(z)$ and in ${\bf E}_{\parallel}$ maintain this %%@
state by limiting ion Lorentz factors and hence photo-electric transition rates.
The ratio,
\begin{eqnarray}
\frac{\rho(h)}{\rho(0)} = \frac{KZ_{s} + 2Z_{h} - Z_{\infty}}
{KZ_{s} + 2\tilde{Z} - Z_{\infty}},    \nonumber
\end{eqnarray}
including ion, proton and electron charge densities
is always very nearly unity at $z < h$.  Here $K$ is the mean number of protons %%@
produced in electromagnetic showers per unit nuclear charge accelerated.  The mean  %%@
charge of nuclei reaching the surface after moving through the electromagnetic shower %%@
region is $Z_{s}$,
and $Z_{\infty}$ is the ion charge at $z_{max}$, where photo-electric transition rates %%@
become negligibly small.
Because the acceleration in this interval is minor compared with that at $z > h$ we can %%@
assume that the effect of this region is to inject ions of mean charge $Z_{h}$ and %%@
Lorentz factor $\gamma _{h}$ into the main acceleration region $z > h$ in which %%@
$\rho(z)$ is modified by photo-electric transitions and possibly by electron-positron %%@
pair creation.  Although the mean value of $Z_{h} - \tilde{Z}$ is small, it is %%@
undoubtedly finite, and the associated reverse-electron flux gives a modest energy %%@
input, $\epsilon _{h}$ , to the polar-cap surface even in the absence of any %%@
whole-surface contribution to photo-ionization at $z > h$.  We refer to Paper III for %%@
its approximate size, $\epsilon _{h} \sim 20$ GeV. 

For a relatively small fraction of pulsars, pair creation by the conversion of %%@
curvature radiation (CR) photons is possible.  But for most ${\bf \Omega}\cdot{\bf B} > %%@
0$ pulsars, it is believed that the source of pairs is the conversion of %%@
outward-directed ICS polar-cap photons (see Hibschman \& Arons 2001, Harding \& %%@
Muslimov 2002).  However, in the case of the ${\bf \Omega}\cdot{\bf B} < 0$ pulsars %%@
considered here, there is an essential difference in that the electron-photon momentum %%@
in the frame of the rotating neutron star is directed inwards.  There is likely to be %%@
some pair creation at low values of $z$ from $\gamma$-rays emitted by the capture of %%@
shower-generated neutrons at
the polar-cap surface, but the rate is difficult to estimate.  Thus pair creation rates %%@
in most ${\bf \Omega}\cdot{\bf B} < 0$ pulsars are likely to be one or two orders of %%@
magnitude smaller than for those with spin direction ${\bf \Omega}\cdot{\bf B} > 0$.  %%@
But even for the latter spin direction, the number of ICS pairs formed per primary %%@
electron can be much smaller than unity (see Harding \& Muslimov 2002, also Fig. 8 of %%@
Hibschman \& Arons).  Thus for the ${\bf \Omega}\cdot{\bf B} < 0$ spin direction, there %%@
is a very real problem in understanding how a density of low-energy secondary %%@
electron-positron pairs adequate for coherent radio emission can be formed.

\section{Conditions for strong turbulence}

There is now a consensus that, in almost all observable radio pulsars, the source of %%@
the coherent emission lies within radii $\eta \sim 10^{1} - 10^{2}$ and its energy is %%@
derived from particle acceleration at the polar cap. There is also a growing consensus %%@
that the formation of strong plasma turbulence is involved (Melrose 2000). The unstable %%@
mode that develops into strong plasma turbulence and transfers energy from  particle %%@
beams to the electromagnetic field must therefore have a high growth rate. This %%@
severely constrains the energy of the beam particles.  The equations of motion for the %%@
system are the Maxwell equations and those for a relativistic particle fluid. In a %%@
strong magnetic field, the mass-to-charge ratio present in the equations is  
$m_{i}\gamma^{3} _{i}q^{-1}_{i}$ for particles of species $i$, where $q_{i}$ is the %%@
charge and $\gamma _{i}$ the Lorentz factor.  The longitudinal effective mass is %%@
$m_{i}\gamma^{3}_{i}$ and  this factor necessarily determines the growth rate of any %%@
unstable mode.

We consider in this paper, specifically, the growth of a quasi-longitudinal Langmuir %%@
mode (Asseo, Pelletier \& Sol 1990) from which a random array of stable Langmuir %%@
solitons may develop (Asseo \& Porzio 2006).  A different longitudinal mode has been %%@
studied by Weatherall (1997, 1998) and may be the source of the nanosecond pulses %%@
observed in certain pulsars (Hankins et al 2003; Soglasnov et al 2004).  (We refer to %%@
the paper of Asseo \& Porzio for an account of the history of plasma turbulence in %%@
relation to pulsar physics.)  The Langmuir mode and solitons have components ${\bf %%@
E_{\parallel}} \neq 0$, ${\bf E_{\perp}} \neq 0$ and ${\bf B}_{t}$ perpendicular to %%@
${\bf E}_{\perp}$ and so can transfer  energy to the radiation field directly (see %%@
Asseo, Pelletier \& Sol 1990).  Hence the direct transfer of energy to the radiation %%@
field is determined only by the longitudinal effective mass and by the departure of the %%@
mode wave-vector from the purely longitudinal state.
Examination of growth rates shows at once that primary electrons or positrons of $\sim %%@
10^{3}$ GeV energy have longitudinal effective masses so large that they are unable to %%@
participate in the mode.  The beams that interact in ${\bf \Omega}\cdot{\bf B} > 0$ %%@
pulsars can be only  secondary electrons and positrons whose velocity distributions %%@
satisfy the relativistic Penrose condition (Buschauer \& Benford 1977).  But ion and %%@
proton beams that have longitudinal effective masses of the same order as secondary %%@
electrons and positrons can be produced in ${\bf \Omega}\cdot{\bf B} < 0$ neutron %%@
stars.  Thus it is anticipated that pulsars of both directions exist and produce very %%@
similar coherent emission.

Beams of cold ions and protons (or, hypothetically, cold electrons and positrons) %%@
permit growth of the quasi-longitudinal Langmuir mode for which the dispersion relation %%@
is,
\begin{eqnarray}
\left(\omega^{2} - k^{2}_{\parallel}\right)\left(1 - \frac{\omega^{*2}_{1}}
{(\omega - k_{\parallel}\beta _{1})^{2}} - \frac{\omega^{*2}_{2}}
{(\omega - k_{\parallel}\beta _{2})^{2}}\right) = k^{2}_{\perp},
\end{eqnarray}
for angular frequency $\omega$, wave-vector ${\bf k}$, and ion and proton (or, electron %%@
and positron) velocities $\beta _{1}$ and $\beta _{2}$, respectively.  The remaining %%@
quantities are
$\omega^{*} _{i} = \gamma^{-3/2}_{i}\omega _{i}$ in which,
\begin{eqnarray}
\omega^{2}_{i} = \frac{4\pi N_{i}q^{2}_{i}}{m_{i}},
\end{eqnarray}
where $q_{i}$, $m_{i}$ and $N_{i}$ are respectively the charge, mass and neutron-star %%@
frame number density of each component $i = 1,2$.  The transverse field components are,
\begin{eqnarray}
E_{\perp}  & =  &  \frac{k_{\perp}E_{\parallel}}{2(k_{\parallel} - \omega)},\nonumber %%@
\\
B_{t} & =  & \frac{k_{\perp}E_{\parallel}}{\omega} - %%@
\frac{k_{\perp}k^{2}_{\parallel}E_{\parallel}}{\omega(k^{2}_{\parallel} - \omega^{2})}.
\end{eqnarray}
The plasma frequency in the rest frame of a component is $\gamma^{-1/2}_{i}\omega _{i}$ %%@
and so, following Asseo et al, a natural wavenumber for which to seek a solution is
defined in terms of the slower-moving component, and is $k_{\parallel} = %%@
2\omega^{*}_{1}\gamma^{2}_{1}$.  With the definition of a new variable $s$ such that,
\begin{eqnarray} 
\omega - k_{\parallel}\beta _{1} = \omega^{*}_{1}(1 + s),
\end{eqnarray}
and,
\begin{eqnarray}
\omega - k_{\parallel} = \omega^{*}_{1}(1 + s) - \frac{k_{\parallel}}{2\gamma^{2}_{1}},
\end{eqnarray}
valid for $\gamma _{1} \gg 1$
equation (8) can be expressed as,
\begin{eqnarray}
1 - \frac{1}{(s + 1)^{2}} - \frac{C}{(s + \mu)^{2}} = %%@
\frac{k^{2}_{\perp}\gamma^{2}_{1}}{k^{2}_{\parallel}s},
\end{eqnarray}
in which $\mu = \gamma^{2}_{1}/\gamma^{2}_{2}$ and 
$C = \omega^{*2}_{2}/\omega^{*2}_{1}$.
In the limit $k_{\perp} \rightarrow 0$, equation (13) is a quartic with two real and %%@
two complex roots, whose properties were investigated in Paper III.  The amplitude  %%@
growth rate was found to be of the order of
${\rm Im}\omega \sim 0.2\omega^{*}_{1}$ and is a slowly varying function of the ratio %%@
$\gamma _{1}/\gamma _{2}$ and, for ions, the proton-ion number density ratio.  But we %%@
have to consider the quasi-longitudinal case of $k_{\perp} \neq 0$ in which the %%@
transverse field components are non-zero. Equation (13) is then less transparent, being %%@
a quintic with three real and two complex roots, but we can use perturbation theory to %%@
estimate the extent to which a non-zero $k_{\perp}$ changes a given complex root.  Let %%@
$s = s_{0} + \delta s$, where $s_{0}$ is a root of the quartic. Then,
\begin{eqnarray}
\delta s\left(\frac{1}{(s_{0} + 1)^{3}} + \frac{C}{(s_{0} + \mu)^{3}}\right) = %%@
\frac{k^{2}_{\perp}\gamma^{2}_{1}}{k^{2}_{\parallel}s_{0}}.
\end{eqnarray}
From equations (10), we see that $|E_{\perp}| = |E_{\parallel}$ for $k_{\perp} = %%@
k_{\parallel}|s|/\gamma^{2}_{1}$.  Then for typical values of $C$, $\mu$ and $s_{0}$,
we have,
\begin{eqnarray} 
|\delta s| \approx \frac{k^{2}_{\perp}\gamma^{2}_{1}}{k^{2}_{\parallel}|s_{0}|}
 = \frac{|s_{0}|}{\gamma^{2}_{1}},
\end{eqnarray}
indicating that for a significant range of $k_{\perp}$, the growth rate does not differ %%@
much from the $k_{\perp} = 0$ value.

The growth rate decreases with distance above the polar cap and so it is appropriate to %%@
obtain the amplitude growth factor $\exp \Lambda$ at a given radius $r = \eta R$ by %%@
integration.  
It is,
\begin{eqnarray}
\Lambda  = {\rm Im}s\left(\frac{-16\pi \alpha _{1}q_{1}R^{2}B\cos\psi}
{Pm_{1}c^{3}\gamma^{3}_{1}}\right)^{1/2} \left(1 - \eta^{-1/2}\right),
\end{eqnarray}
assuming a dipole field and constant $\gamma _{1}$. Here, the ion (or electron) %%@
component number density in the frame of the rotating neutron star is a fraction %%@
$\alpha _{1}$ of the Goldreich-Julian number density. For ${\rm Im}s = 0.2$ and %%@
$\cos\psi = -1$, the factor is,
\begin{eqnarray}
\Lambda = 2.4\times 10^{5}
\left(\frac{B_{12}q_{1}m\alpha _{1}}{Pem_{1}\gamma^{3}_{1}}\right)^{1/2}
\left(1 - \eta^{-1/2}\right),
\end{eqnarray}
where $m$ is the electron rest mass.
We can see that the growth factor increases most rapidly at small $\eta$ and is %%@
substantially developed at $\eta = 2$.  The only particle property that it depends on %%@
is the charge to longitudinal effective mass ratio.  A minimum value necessary for the %%@
development of nonlinearity is presumably given by the size of particle number %%@
fluctuations within a cubic-wavelength volume at the Goldreich-Julian number density.  %%@
This indicates a minimum $\Lambda \sim 20$.  We assume $\Lambda = 30$.
For a typical pulsar with $PB^{-1}_{12} = 1$, equation (17) then requires  $\gamma _{1} %%@
< 176\alpha^{1/3}_{1}$ for electrons or $\gamma _{1} < 11\alpha^{1/3}_{1}$ for ion %%@
beams at $\eta = 2$ increasing to $\gamma _{1} < 311\alpha^{1/3}_{1}$ and
$\gamma _{1} < 20\alpha^{1/3}_{1}$, respectively, at $\eta = 10$.
Development of turbulence or of stable solitons is then possible for either electrons %%@
or ions although for electrons, the cold beam hypothesis is clearly unrealistic and the %%@
distributions of electron and positron velocities must satisfy the Penrose condition.

Quasi-longitudinal modes transfer energy directly to the radiation field at angular  %%@
frequencies of the order of $\omega _{0} = 2\gamma^{1/2}_{1}\omega _{1}$.  The actual %%@
maximum in the intensity distribution presumably depends on the extent to which %%@
developed turbulence moves energy toward higher wavenumbers.  This is unknown except %%@
that some examples of soliton formation considered by Asseo \& Porzio have maxima at %%@
$\omega \approx 1.5\omega _{0}$. As a function of $\eta$ and for $\cos\psi = -1$, the %%@
frequencies $\nu _{0} = \omega _{0}/2\pi$ are,
\begin{eqnarray}
\nu _{0} &  = & 4.8\times 10^{9}\left(\frac{B_{12}\alpha _{1}}{P}\right)^{1/2}
\left(\frac{\gamma _{1}}{\eta^{3}}\right)^{1/2}     \nonumber  \\
\nu _{0} & = & 1.1\times 10^{8}
\left(\frac{B_{12}Z_{\infty}\alpha _{1}}{AP}\right)^{1/2}
\left(\frac{\gamma _{1}}{\eta^{3}}\right)^{1/2}
\hspace{3mm}{\rm Hz} 
\end{eqnarray}
respectively, for electrons and for ions of mass number $A$ and charge $Z_{\infty}$. %%@
These values differ by two orders of magnitude and suggest the existence of two classes %%@
of pulsar with very different radiation spectra though with individual pulses of %%@
similar form. There have been relatively few measurements of flux density  below 400 %%@
MHz but those of Deshpande \& Radhakrishnan (1992) and of Malofeev et al (1994) both %%@
confirm that low-frequency spectral cut-offs are usually below $100$ MHz and in some %%@
cases, below $50$ MHz.

There have been many calculations of pair formation densities above polar caps, but of %%@
particular interest is the paper of Harding \& Muslimov (2011) who have studied the %%@
effect of a specific type of field, that of an off-set dipole.  Pair formation %%@
densities, assuming outward moving ICS photons to be the source, are rapidly increasing %%@
functions of the degree of off-set. The most extreme case in which pair formation might %%@
be thought doubtful is that of PSR J2144-3933 which has a period $P = 8.51$ s and whose %%@
observers (Young, Manchester \& Johnston 1999) commented specifically on the problem.    %%@
The value of its parameter
$BP^{-2} = 2.87\times 10^{10}$ G s$^{-2}$ is almost an order of magnitude
smaller than the apparent cut-off value of $2.2\times 10^{11}$ G s$^{-2}$ in the %%@
distribution of that quantity obtained from the ATNF pulsar catalogue (Manchester et al %%@
2005) representing a death-line in the $P$-$\dot{P}$ distribution below which pair %%@
creation is not possible. It is old ($2.7\times 10^{8}$ yr) with a relatively low mean %%@
flux density at 400 MHz, and may be observable only as a consequence of close proximity %%@
(170 pc).  Although the maximum acceleration potential given by equation (5) is very %%@
small, $V_{max} = 36$ GeV, ICS pair formation is possible in principle although a very %%@
high degree of off-set would be required (see Figs. 7 - 9 of Harding \& Muslimov 2011). %%@
However, given a spin direction ${\bf \Omega}\cdot{\bf B} < 0$, accelerated ions would %%@
have Lorentz factors appropriate for growth of the quasi-longitudinal mode for any %%@
field geometry.  But the value of $\nu _{0}$ given by equations (18) is rather small in %%@
this case, even assuming that the radiation field is largely decoupled from the %%@
magnetosphere at quite low altitudes $\eta \sim 2$.  (For this spin direction, the pair %%@
multiplicities produced by ICS photons would probably be one or two orders of magnitude %%@
smaller than for ${\bf \Omega}\cdot{\bf B} > 0$ because the momentum of the %%@
electron-photon system is inward directed.) But more extensive observations at low %%@
frequencies may reveal whether or not J2144-3933 is unique or merely a nearby example %%@
of an otherwise unobserved population.

\section{Instability and proton production}

Owing to our late recognition that photo-electric transitions have a very significant %%@
effect on particle acceleration under SCLF boundary conditions, it is necessary to %%@
revise the estimates of the reverse-electron energy flux that were used in Papers I and %%@
II.  Photo-electric transitions and the consequent reverse-electron flux produce %%@
changes in mean nuclear and ion charge that we can summarize here as follows. The %%@
initial nuclear charge is $Z$ (canonical value $Z = 26$) but formation and decay of the %%@
giant dipole state in electromagnetic showers reduces it to $Z_{s}$ at the top of the %%@
neutron-star atmosphere.  We assume local thermodynamic equilibrium (LTE) in this %%@
region with ion charge  $\tilde{Z}$.  Then ions with charge $Z_{h}$ are injected into %%@
the main acceleration region at $z = h$. Typically, $Z_{h}\approx \tilde{Z} + 1$ owing %%@
to interaction with polar-cap blackbody photons.  Then further acceleration produces %%@
more extensive ionization to $Z_{\infty} \leq Z_{s}$ by the whole-surface blackbody %%@
field with Z(z) as the altitude-dependent ion charge during this process. 

In papers I-III we defined $KZ_{s}$ to be the number of protons produced per ion of %%@
nuclear charge $Z_{s}$ accelerated.  The initial problem treated here is of %%@
acceleration given by equations (2) - (4) with $\rho(z)$ independent of time and of %%@
position ${\bf u}$.  Then,
\begin{eqnarray}
\frac{\rho(z)}{\rho(h)} = \frac{KZ_{s} + 2Z(z) - Z_{\infty}}{KZ_{s} + 2Z_{h} - %%@
Z_{\infty}},  \nonumber
\end{eqnarray}
and the total photo-ionization at altitude $z$ is related to the potential at that %%@
point by,
\begin{eqnarray}
\frac{2Z(z) - 2Z_{h}}{KZ_{s} + 2Z_{h} - Z_{\infty}} = \frac{\kappa(\eta^{-3}_{h} - %%@
\eta^{-3})}
{1 - \kappa \eta^{-3}_{h}}\left(1 - \frac{V({\bf u},z)}{V_{max}({\bf u},z)}\right),
\end{eqnarray}
where $\eta _{h} = 1 + h/R$.
This result takes into account protons, ions and the reverse flux of photo-electrons, %%@
but not CR electron-positron pair creation or the conversion of ICS photons.

Photo-electric transition rates are rapidly increasing functions of the ion Lorentz %%@
factor $\gamma$.  The reverse-electron flux they produce leads to a charge density  %%@
that increases with altitude, so reducing the potential given by equation (2).  Thus we %%@
might expect a self-consistent set of functions $V(z,{\bf u})$, $\rho(z,{\bf u})$ and %%@
$\gamma(z,{\bf u})$ that are time-independent. In order to investigate this %%@
possibility, it is necessary to calculate the reverse-electron energy flux taking into %%@
account the fact that photo-electric transitions reduce the potential energy $V = - %%@
e\Phi$ derived from equation (2) to levels below $V_{max}$.  Broadly, we expect that %%@
$\gamma(z)$ will increase from its initial value $\gamma _{h}$
but that $V({\bf u},\infty) < V_{max}({\bf u},\infty)$, and our initial attempt to %%@
model this is to assume that $V({\bf u},z) = V_{max}({\bf u},z)$ 
until a cut-off $V_{c}$ is reached.  

With the procedures and assumptions of Paper III, we calculate the mean  %%@
reverse-electron energy $\epsilon _{s}$ per ion resulting from interaction with %%@
whole-surface blackbody photons, and the mean ion charge $Z_{\infty}$.
Transition rates are obtained to an altitude $z_{max} = 3R$, beyond which they become %%@
small and equation (2) predicts negligible further acceleration.  The polar-cap %%@
temperature is treated as distinct from the general surface temperature of the star and %%@
its value is equal to that generated by the reverse-electron flux which produces %%@
protons at the Goldreich-Julian current density.
A flux exceeding this produces an accumulation of protons at the top of the
neutron-star atmosphere which, until exhausted, reduces the ion component of the %%@
current density to zero.  We adopt equation (33) of Paper I, with $\cos\psi = -1$ and %%@
rotation period $P = 1$ s, and a typical proton production rate per unit shower energy %%@
of $W_{p}mc^{2} = 2\times 10^{-4}$, that is, one proton per $2.5$ GeV shower energy.  %%@
This quantity is a slowly varying function of $B$ and $Z$ and has been estimated in %%@
Papers I and II.   Then $(\epsilon _{h} + \epsilon _{s})W_{p} = KZ_{s}$ with $\epsilon %%@
_{h} \sim 20$ GeV being the contribution of polar-cap blackbody photons.
The temperatures are $T_{pc} = 0.74\times 10^{6}$ K, $0.82\times 10^{6}$ K and %%@
$1.10\times 10^{6}$ K respectively for $B = 1.0\times 10^{12}$, $3.0\times 10^{12}$ and %%@
$10^{13}$ G. These have negligible effect on the acceleration but determine the initial %%@
ion charge $\tilde{Z}$.
Actual values of $Z_{s}$ are unknown {\it ab initio} but are given by $Z_{s} = Z/(1 + %%@
\bar{K})$, where $\bar{K}$ is the time-average of $K$ at a specific position ${\bf u}$ %%@
on the polar cap and $Z$ is here the pre-shower nuclear charge.  Hence we assume %%@
arbitrary values $Z_{s} = 10$ or $20$ with mass numbers $A = 20$ or $40$ and note that %%@
ions with very small $Z_{s}$ tend to be completely ionized by the polar-cap or %%@
whole-surface temperature at the start of acceleration and so produce no %%@
reverse-electron flux. The initial ion Lorentz factor is fixed as $\gamma _{h} = 5$ at %%@
an altitude $z = h = 0.05R$. The initial ion charges at $z = h$ are: for $Z_{s} = 10$; %%@
$Z_{h} = 8,6$ or $5$ and for $Z_{s} = 20$; $Z_{h} = 15,12$ or $9$ respectively at $B = %%@
1.0\times 10^{12}$, $3.0\times 10^{12}$ G or $10^{13}$ G..

Values of $\epsilon _{s}$ and $Z_{\infty}$ obtained on this basis are given in Table 1.
With the approximations and methods of Paper III in mind, it is obvious that the %%@
absolute values of $\epsilon _{s}$ do not have the accuracy that the number of figures %%@
given might indicate.  But comparison of columns for different values of $T_{s}$ shows %%@
that whole-surface photons can generate very high transition rates not only at large %%@
$V_{c}$ for $T_{s} = 10^{5}$ K, but for progressively lower $V_{c}$ as whole-surface %%@
temperature increases.  By combining the results of Table 1, which are estimates of %%@
$\epsilon _{s}(V)$ and $Z_{\infty}(V)$, with equation (19) it should be possible, given %%@
$Z_{h}$, to solve for $V({\bf u},\infty)$.  It can be seen by inspection of the %%@
equation, with reference to values of $K(V)$ found from the Table, that there is always %%@
a solution, the value of $V$ depending principally on $T_{s}$ and to a lesser extent on %%@
$B$.  Values $V \ll V_{max}$ occur at high $T_{s}$ and small $B$.

Basically, incomplete ionization at $z = h$ means that there will be further ionization %%@
at $z > h$ unless $T_{s}$ or $V_{max}$, or both, are too small for it to be possible.
In this latter case, relevant for old pulsars, there remains the polar-cap contribution %%@
$\epsilon _{h}$ giving a small factor $K < 1$ and a solution $V({\bf u},z) = %%@
V_{max}({\bf u},z)$.  Such a solution would be time-independent and of uniform $\rho$
over the polar cap, with both ion and proton components.  But it would not lead to %%@
radio emission unless $V_{max}$ were small enough to give an adequate growth rate for %%@
the Langmuir mode, as described in Section 3.  The general cut-off $BP^{-2} = 2.2\times %%@
10^{11}$ G s$^{-2}$ obtained from the ATNF pulsar catalogue gives $V_{max}({\bf %%@
u},\infty) = 275(1 - u^{2}/u^{2}_{0})$ GeV.  Ion Lorentz factors $\gamma \sim 20$ and
proton energies $\sim 60$ GeV are possible on flux lines originating near the edge of %%@
the polar cap.  However, in the more general case, large values of $\epsilon _{s}$ in %%@
the uniform time-independent model lead to $K \gg 1$ and hence to instability, as shown %%@
in Section 4.3 of Paper I.  The actual state of the polar-cap surface is therefore %%@
time-dependent.

\section{Qualitative modelling of the polar cap}

The basic principle is that the state of any element $\delta {\bf u}$ of surface at any %%@
instant is a function only of its past time-distribution of reverse-electron energy %%@
flux. The flux of accelerated protons $J^{p}({\bf u},t)$ is given by,
\begin{eqnarray}
J^{p}({\bf u},t) + \tilde{J}^{p}({\bf u},t) = \int^{t}_{-\infty}f_{p}(t - %%@
t^{\prime})K({\bf u},t^{\prime})J^{z}({\bf u},t^{\prime}),
\end{eqnarray}
in terms of the ion flux $J^{z}$. The quantity $\tilde{J}^{p} = 0$ within intervals for %%@
which $J^{p}$ does not exceed the Goldreich-Julian current density $\rho _{GJ}(0)c$. %%@
Most protons are produced in the vicinity of the shower maximum whose depth is only a %%@
slowly varying function of the primary electron energy.
Their motion toward the surface is better described by a drift velocity, rather than %%@
diffusion, in most of the LTE atmosphere so that the distribution of arrival times can %%@
be modelled as $f_{p}(t - t^{\prime}) = \delta(t - t^{\prime} - \tau _{p})$.  We refer %%@
to Section 5 of Paper II for further details.  Following Paper III, equation (20) %%@
recognizes that $K$ is a function of ${\bf u}$ and $t$ owing to its $V-$dependence, and %%@
can vary on time-scales of the order of $\tau _{p}$.
The protons are preferentially accelerated, but if $J^{p}$ reaches the Goldreich-Julian %%@
flux, as can be the case if $K \gg 1$, a thin proton atmosphere forms at the top of the %%@
ion atmosphere at a rate given by $\tilde{J}^{p}$ and the local ion flux falls to zero %%@
until it is exhausted.

As a first approximation, equation (20) shows that the condition of an element of %%@
polar-cap area can be represented by two possible states: ion emission with duration %%@
$\tau _{p}$ or proton emission of duration $\tau _{gap} \approx \bar{K}\tau _{p}$, %%@
where $\bar{K}$ is here the time-average over the interval $\tau _{p}$.  Values %%@
$\bar{K} \gg 1$ ensure that proton production is so large that an atmosphere of protons %%@
forms at the top of the LTE atmosphere of ions.  Then the ion phase ends and the proton %%@
phase is maintained until the proton component of the atmosphere is exhausted.  Areas %%@
in the ion phase necessarily move with time and so can be seen as moving subpulses %%@
within the time-averaged radio-pulse profile.  The motion could be chaotic or, as %%@
proposed in Paper II, organized so as to give regular subpulse drift. 

Pair creation by curvature radiation (CR) photons was referred to briefly at the end of %%@
Section 2 but has been otherwise ignored here.  The necessary conditions are thought to %%@
be present in a relatively small fraction of the observed population (Harding \& %%@
Muslimov 2002) but we shall review briefly its effects in the
${\bf \Omega}\cdot {\bf B} < 0$ case with SCLF boundary conditions. Self-sustaining %%@
pair creation requires adequate conversion probabilities for both outward and %%@
inward-directed CR photons at $z < z_{max}$.  Thus pair creation is not uniform over %%@
the whole cross-sectional area $\pi u^{2}_{0}(z)$ of the open flux tube,  but depends %%@
on a combination of $V$ and of flux-line curvature.  For the suitable values of ${\bf %%@
u}$, the equation analogous to (19), obtained by comparison of current densities at $z %%@
= 0$ and $z = z_{max}$, can be most compactly expressed in terms of the particle flux %%@
$\phi$ on the relevant flux lines,
\begin{eqnarray}
\frac{\phi _{p} + |\phi _{e}|}{\phi _{p} - |\phi _{e}|} = 1 + \frac{\kappa}{1 - \kappa}
  \left(1 - \frac{V({\bf u},\infty)}{V_{max}({\bf u},\infty)}\right),
\end{eqnarray} 
in which $\phi _{e}$ is the flux either of electrons or positrons, and $\phi _{p}$ is %%@
the proton flux.  This assumes, as is almost certainly the case, that the %%@
reverse-electron energy flux produces an excess of protons at the top of the LTE %%@
atmosphere.  There is then no ion flow except on flux lines that do not satisfy the CR %%@
criteria such as those near $u_{0}$ on which $V_{max}$ is too small. 

\begin{table}
\caption{Values of the mean reverse-electron energy per ion ($\epsilon _{s}$) at the %%@
polar-cap surface in units of GeV, and of the mean ion charge at the end of %%@
photo-ionization ($Z_{\infty}$) have been calculated.  The magnetic flux density is in %%@
units of $10^{12}$ G and the acceleration potential cut-off $V_{c}$ is in units of GeV.  %%@
Arbitrary values $Z_{s} = 10$ or $20$ have been assumed for the mean surface nuclear %%@
charge. We refer to Section 4 for discussion of $Z_{s}$ and other parameters on which %%@
the calculated values are dependent. The final three pairs of columns give the values %%@
of $\epsilon _{s}$ and of $Z_{\infty}$ for whole surface temperatures $T_{s} = 1,2$ and %%@
$4 \times 10^{5}$ K. The rotation period is $P = 1$ s. Blank spaces denote energies %%@
that are negligibly small.}

\begin{tabular}{@{}rrrrrrrrr@{}}
\hline
$B_{12}$ & $Z_{s}$ & $V_{c}$ & $\epsilon _{s}$ & $Z_{\infty}$ & $\epsilon$ & %%@
$Z_{\infty}$
& $\epsilon$ & $Z_{\infty}$  \\
 &  & GeV & $T_{s5}$ &$= 1$ &  & $2$ &  & $4$   \\
\hline

1.0 & 10 & 1000 & 1389 & 9.9 & 650 & 10.0 & 255 & 10.0  \\
	&    &  500 & 253 & 8.5 & 650 & 10.0 & 255 & 10.0  \\
	&    &  250 &   4 & 8.0 & 392 & 9.6 & 255 & 10.0  \\
	&    &  125 &     &     & 38  & 8.3 & 228 & 10.0  \\
	&    &   80 &     &     & 12 & 8.1 & 128 & 9.6  \\
	&    &   50 &     &     &  2 & 8.0 & 67 & 9.2  \\
	
	& 20 & 1000 & 2196 & 18.0 & 2044 & 20.0 & 803 & 20.0  \\
	&    &  500 & 328 & 15.7 & 1783 & 19.7 & 803 & 20.0  \\
	&    &  250 &  5 & 15.0 & 640 & 17.5 & 802  & 20.0  \\
	&    &  125 &    &      &  56 & 15.4 & 395 & 18.4 \\
	&    &   80 &    &      &  14 & 15.2 & 241 & 17.9  \\
	&    &   50 &    &      &   2 & 15.0 & 136 & 17.3  \\
	
3.0 & 10 & 2000 & 6438 & 9.3  & 4264  & 10.0 & 1607 & 10.0  \\
	&    &  1000 & 138 & 6.1 &  3767 & 10.0  & 1607 & 10.0  \\
	&    &  500 & 36 & 6.1 & 581 & 7.2 & 1524 & 10.0  \\
	&    & 250  & 7 & 6.0 & 244 & 7.0 & 734 & 9.3  \\
	&    &  125 &  &  & 64 & 6.5 & 373 & 8.9  \\
	&    &  80 &   &  & 16 & 6.2 & 225 & 8.5  \\
	&    &  50 &   &  & 2 & 6.0 & 119 & 7.9  \\
	
	& 20 & 2000 & 10580 & 17.4 & 9196 & 20.0 & 3760 & 20.0  \\
	&    &  1000 & 104 & 12.1 & 6677 & 18.9 & 3760 & 20.0  \\
	&    &  500 & 27 & 12.1 & 647 & 13.3 & 3346 & 19.8  \\
	&    &  250 & 3 & 12.0 & 225 & 12.9 & 1348 & 17.7  \\
	&    &  125 &   &    & 35 & 12.3 & 604 & 16.5  \\
	&    &   80 &     &     & 5 & 12.1 & 319 & 15.4  \\
	&    &   50 &    &      &    &      & 124 & 14.0  \\

10.0 & 10 & 2000 & 85 & 5.0 & 3114 & 6.6 & 8521 & 10.0  \\
	 &    &  1000 & 47 & 5.0 & 631 & 5.7 & 3578 & 8.9  \\
	 &    &  500 & 12 & 5.0 & 345 & 5.7 & 1724 & 8.6  \\
	 &    &  250 & 1 & 5.0 & 98 & 5.4 & 814 & 8.1  \\
	 &    &  125 &     &     &  10 & 5.1 & 314 & 7.3  \\
	 &    &   80 &     &     &   1 & 5.0 & 144 & 6.5  \\
	 &    &   50 &     &     &     &     & 47 & 5.7  \\

	 & 20 & 2000 & 137 & 9.1 & 3061 & 10.6 & 19778 & 20.0  \\
	 &    & 1000 & 58 & 9.1 & 1055 & 10.1 & 6523 & 16.1  \\
	 &    &  500 & 10 & 9.0 & 456 & 9.9 & 3269 & 15.6  \\
	 &    &  250 &   1 & 9.0 & 90 & 9.3 & 1388 & 14.3  \\
	 &    &  125 &     &      & 5 & 9.0 & 440 & 12.1  \\
	 &    &   80 &     &      &     &      & 155 & 10.6  \\
	 &    &   50 &     &      &     &      & 31 & 9.5  \\
\hline
\end{tabular}
\end{table}

Quantitative model construction is not attempted here.  Owing to the uncertainties in %%@
significant parameters that are listed in Section 6, quantitative testing of a model %%@
against observed data does not appear immediately profitable.  The purpose of this %%@
paper is less ambitious.  It is an attempt to show that the condition of the polar-cap %%@
surface provides a physical basis for nulls, mode-switches and subpulse drift. To do %%@
this, we describe qualitatively some of the possible polar-cap states.

Assume, initially, that CR pair production as described above is possible over some %%@
limited area of the polar cap.  Its boundary can be defined as that within which proton %%@
production by the reverse-electrons of CR pairs exceeds the Goldreich-Julian rate so %%@
that a proton atmosphere forms and grows in density.  Between this boundary and ${\bf %%@
u_{0}}$ there is an annular strip within which ion acceleration can occur and proton %%@
formation is governed by equation (20).  (Any contribution from CR reverse-electrons is %%@
neglected here.)  Solutions of equation (20) then exist in which zones of ion phase %%@
circulate around the CR region.
Equation (20) is local in ${\bf u}$ and so cannot itself determine the circulation %%@
sense or velocity.  Suppose that there are $n$ ion zones.  Then in the usual notation %%@
by which subpulse drift is described, the circulation time is
$\hat{P}_{3} = nP_{3}$, where $P_{3}$ is the band separation and in the model is given %%@
by $P_{3} = \tau _{gap} + \tau _{p} = (K + 1)\tau _{p}$.
Association with subpulse drift would require either secondary pair creation on the ion %%@
phase flux lines or the presence of proton and ion components in the current density %%@
near to ${\bf u_{0}}$ so that energies are low enough for Langmuir-mode growth.  A %%@
further consideration is that excess proton production within the CR area must be %%@
accompanied by lateral diffusion on the polar-cap surface, perpendicular to ${\bf B}$.  %%@
We have chosen the example of a clearly defined annular region but there is no reason %%@
why organized motion of this kind should not occur more generally and in the absence of %%@
a CR component.

 A further possibility is that the active CR region could be switched off by changes in %%@
the potential inside it which is also dependent on the state of the remaining parts of %%@
the polar cap.   In general, in a less-organized state, there is no reason why either %%@
the whole polar cap or large parts of it should not be in the proton phase for %%@
intervals of time of the order of $\tau _{gap}$ and, in the absence of CR pair %%@
production, there would be no possibility of Langmuir mode growth.  Thus nulls are a %%@
natural phenomenon within the model.
Obviously, there are many possible polar-cap states, their relevance to widely-observed %%@
phenomena being limited only by the requirement that they should not be too dependent %%@
on neutron-star parameters having critical or particular values.

Mode-changes have been studied extensively in a small number of pulsars, some of which %%@
also exhibit subpulse drift.  The paper of Bartel et al (1982) summarizes the main %%@
characteristics of this phenomenon.  These authors emphasize that the spectral and %%@
polarization changes that occur on mode-switching are consistent with a change in the
optical depth of the source region within the magnetosphere and that the sources of the %%@
two or more modes would be on different sets of flux lines. (They also suggest that the %%@
explanation of mode-switching lies in the polar-cap surface, though in the context of %%@
the Ruderman \& Sutherland model.)  Switching of emission between different sets of %%@
flux lines occurs naturally in our qualitative model which is sufficiently diverse to %%@
accommodate the characteristics of mode-switching that are observed, including changes %%@
in subpulse drift rates. 

The qualitative model is applicable to subpulse drift and to short-duration nulls or %%@
mode changes and there is no difficulty in understanding how non-randomness or %%@
quasi-periodicity, such as that reported by Rankin \& Wright (2008) and Redman \& %%@
Rankin (2009) might occur naturally in such a system.  The time-scales are necessarily %%@
short, perhaps of the order of $\tau _{gap}$, but there is also the possibility of a %%@
medium time-scale instability in the atomic number $Z_{s}$ of surface nuclei, described %%@
in Section 4 of Paper II.  This can result in values of $Z_{s}$ that are too small to %%@
produce any reverse-electron flux because the atoms are completely ionized in the LTE %%@
atmosphere.  Having the largest charge-to-mass ratio, apart from protons, they are in %%@
equilibrium at the top of the atmosphere and so are preferentially accelerated through %%@
the full potential difference $V({\bf u},\infty)$ with no possibility of
Langmuir-mode growth if CR pair creation is not possible.  A broad measure of the %%@
time-scale for this instability is provided by the time in which a depth of surface %%@
nuclei equivalent to one radiation length is removed from the polar cap at the %%@
Goldreich-Julian current density.  From equations (4) and (5) of Paper I, this is given %%@
by,
\begin{eqnarray}
t_{rl} = 2.1\times %%@
10^{5}\left(\frac{-P\sec\psi}{ZB_{12}\ln(12Z^{1/2}B_{12}^{-1/2})}\right) \hspace{2mm}
{\rm s}.
\end{eqnarray}
The distribution of $Z_{s}$ over the polar cap at any instant need not, and probably %%@
will not, be uniform.  Thus complex distributions of potential and of accelerated beam %%@
composition are to be expected.

\section{Conclusions and uncertainties}

Both neutron-star spin directions, defined by the sign of ${\bf \Omega}\cdot{\bf B}$ at %%@
the polar caps, presumably exist and it is one of the purposes of this paper to %%@
question how, if at all, each sign contributes to the observed radio pulsar population.
Previous papers (I - III) have been directed toward the ${\bf \Omega}\cdot{\bf B} < 0$ %%@
case because it has positive polar-cap corotational charge density so that the flux of %%@
accelerated particles can have ionic, proton and positron components.  There appears to %%@
have been some reluctance in the published literature to consider in any detail the %%@
physics of the polar cap for this spin direction.  But Papers I - III have attempted to %%@
show that such considerations are essential and that the radio emission characteristics %%@
are not solely determined by electrodynamics, as may be so for ${\bf \Omega}\cdot{\bf %%@
B} > 0$.

Weltevrede, Edwards \& Stappers (2006) detected subpulse drift in almost one half of a %%@
sample of $187$ pulsars selected only by signal-to-noise ratio.  Nulls and mode changes %%@
are less frequently observed.  We propose that results obtained in Papers I - III and %%@
in the present paper lead to the idea that these phenomena are all related to the %%@
physics of the polar cap in ${\bf \Omega}\cdot{\bf B} < 0$ pulsars and that the %%@
opposite spin direction, with only electrons as primary accelerated particles, is then %%@
observed as the population showing subpulse modulation, perhaps as a consequence of %%@
plasma turbulence, but none of the above phenomena.  The problem with polar-cap physics %%@
is that a real condensed-matter system can be extremely complex and, as we noted in %%@
Section 1, does not always lend itself to the construction of simple models (see also %%@
Section 2 of Paper II).  A further difficulty is that some of the parameters concerned %%@
are either unknown, such as the atomic number of surface nuclei, or can vary over %%@
several orders of magnitude.

Calculated values in Table 1 are limited to $B = 10^{12} - 10^{13}$ G which is %%@
representative of pulsars in the nulls tabulation of Wang et al (2007) and, for %%@
example, of those pulsars with drifting subpulses investigated by Gil et al (2008).    %%@
Photo-electric cross-sections obtained in Paper III are not valid at $B < 10^{12}$ G in %%@
which region interpolation between zero-field cross-sections and those at $10^{12}$ G
would be necessary.  At higher fields, $B > B_{c} = 4.41 \times 10^{13}$ G, approximate %%@
cross-sections for bremsstrahlung and pair creation have been found (Jones 2010b) and %%@
electromagnetic shower development investigated.  It was confirmed that the value of %%@
the proton production parameter $W_{p}$ decreases only slowly at $B > B_{c}$ but also %%@
that the Landau-Pomeranchuk-Migdal effect, known to exist in zero field, is significant %%@
in the neutron-star atmosphere, and has the effect of reducing high-energy %%@
bremsstrahlung and pair creation cross-sections and therefore of increasing shower %%@
depth and hence the proton diffusion-drift time $\tau _{p}$.

The whole-surface temperature $T_{s}$ is a further important parameter that is not well %%@
known.  Paper III did not incorporate general-relativistic corrections that should %%@
strictly have been made in transforming the blackbody radiation field to the ion rest %%@
frame.  But for most transitions, the radiation field is better approximated by that of %%@
the neutron-star surface proper frame rather than by $T_{s}^{\infty}$, the temperature %%@
inferred by a distant observer.  Since $T^{\infty}_{s}\approx 0.8T_{s}$ for the %%@
neutron-star mass and radius assumed here in Section 2, the temperatures used in Table %%@
1 are well below those that can presently be observed.  Yakovlev \& Pethick (2004) list %%@
only a small number of young pulsars whose blackbody temperatures have been measured. %%@
These are $T^{\infty}_{s} > 5\times 10^{5}$ K.  All model predictions reviewed by these %%@
authors show $T^{\infty}_{s}$ decreasing very rapidly at age $> 10^{6}$ yr.  But on the %%@
other hand, temperatures below $T_{s} = 10^{5}$ K, the lowest used in Table 1, could be %%@
maintained by a very small and obscure level of dissipation within the star.  Isolated %%@
neutron stars (INS) certainly have higher temperatures $T^{\infty}_{s}\sim 10^{6}$ K 
(see, for example, Kaplan \& van Kerkwijk 2009).  But they are few in number and their %%@
radio beams (if they exist) would be narrow compared with the $4\pi$ observability of %%@
unpulsed blackbody radiation. Given this uncertainty, we do not regard them as part of %%@
the population considered here.

The energy of the coherent radio emission must presumably be transferred from particle %%@
beams in high magnetic field regions near the polar caps.  Whatever the coherent %%@
emission mechanism is eventually established to be, provided it is not curvature %%@
radiation, the criteria in Section 3 for growth of Langmuir modes must be relevant.  %%@
This paper does not attempt to judge whether low-energy secondary electrons or small %%@
Lorentz factor ions form the particle beams producing radio emission in any particular %%@
group, but asserts that both possibilities exist for ${\bf \Omega}\cdot{\bf B} < 0$ %%@
pulsars.  It follows that evolution of a pulsar to old age may include long intervals, %%@
perhaps of the order of $10^{6}$ yr, in which pair creation is not possible because %%@
$V_{max}$ is too small but ion Lorentz factors remain too large to give adequate %%@
Langmuir-mode growth rates.
Because $V_{max}({\bf u},\infty) \propto u^{2}_{0} - u^{2}$, at least approximately, %%@
the highest growth rates will be on flux lines with $u$ near $u_{0}$.  This is entirely %%@
consistent with the conal pattern of emission in the pulsar morphology developed by %%@
Rankin (1983) .

Pulsars exhibit nulls with mean duration as short as several rotation periods or, for %%@
those described as intermittent pulsars, times of the order of $10$ d.
PSR 1931+24 is an example of the latter (Kramer et al 2006), but long-duration nulls %%@
have now also been observed in J1832+0029 (see Lyne 2009) and in J1841-0500 (Camilo et %%@
al 2012). Such null durations allow separate measurement of the spin-down rate in both %%@
on and off-states of emission.  In each case, the off-state spin-down rate is about %%@
half that of the on-state, indicating some substantial difference in the condition of %%@
the magnetosphere.  Obviously, it is not known if this is also true for short-duration %%@
nulls but for ${\bf \Omega}\cdot{\bf B} < 0$ pulsars in our model of the polar cap, it  %%@
occurs naturally owing to the change in the extent to which the electromagnetic fields
near and beyond the light cylinder
are loaded and the total momentum density modified by particle emission.  If the whole %%@
polar cap were in the proton phase, as discussed in Section 5, there would be no %%@
significant pair creation (assuming no CR pair production) and, of course, no %%@
possibility of Langmuir-mode growth.  However, if creation of low-energy secondary %%@
pairs were possible, the flux of particles to be accelerated at or beyond the light %%@
cylinder could greatly exceed the Goldreich-Julian value.  Even though we lack a %%@
comprehensive understanding of acceleration at $R_{LC}$ and beyond, it is entirely %%@
plausible that the changes in particle number density inherent in our model would lead %%@
to the observed changes in spin-down torque.

Interesting reviews of pulsars with large nulling fractions and of the Rotating Radio %%@
Transients (RRAT) have been given recently by Burke-Spolaor \& Bailes (2010), Keane et %%@
al (2011) and by Keane \& McLaughlin (2011).  In particular, pulsars are shown in the %%@
$g - P$ plane, where $g$ is here the fraction of periods in which a pulse is detected.  %%@
The distribution appears roughly uniform in $\log g$ within the interval $10^{-4} < g < %%@
10^{-1}$.  The existence of these objects is not inconsistent with the complexity of %%@
possible polar-cap states of the ${\bf \Omega}\cdot{\bf B} < 0$ pulsars described here
given the time-scale $t_{rl}$ defined by equation (22).  But Keane \& McLaughlin note %%@
that there is a scarcity of objects whose behaviour is governed by time-scales that are %%@
longer than the above but shorter than the very long time-scales
($\sim 10^{6} - 10^{7}$ s) associated with the intermittent pulsars such as PSR %%@
1931+24.
Very long time-scales much exceeding those given by equation (22) still remain a %%@
problem whose explanation may be more subtle than those attempted in this paper or may %%@
lie in quite different considerations.

A simple explanation for the RRAT has been proposed by Weltevrede et al (2006).  These %%@
authors compared RRAT pulses with the infrequent very large-amplitude pulses seen in %%@
PSR 0656+14.  They note that if 0656+14 were as distant in the galaxy as most of the %%@
RRAT, its observed emission would appear to be that of the typical RRAT.  But this may %%@
be inconsistent with the periodicities in individual RRAT pulse arrival times found %%@
recently by Palliyaguru et al (2011).  These are mostly of the order of $10^{4}$ s
but a minority are much longer, of the order of $10^{6} - 10^{7}$ s.  If these complex %%@
sets of periodicities are established and are not loose quasi-periodicities, they %%@
represent a very real problem for the model of the polar cap described here.  It is %%@
difficult to see how they can be present in a truly isolated neutron star, with either %%@
spin direction, not interacting with any other periodic system.  Quasi-periodicities %%@
might be understandable in terms of natural phenomena at the condensed-matter surface, %%@
possibly connected with the diffusion of low-$Z_{s}$ nuclei perpendicular to ${\bf B}$ %%@
in the vicinity of the polar cap or with the two sections of the open magnetosphere %%@
described in Section 2.  But well-defined periodicities would remain a problem.

\appendix

\bsp

\label{lastpage}


\begin{thebibliography}{99}

\bibitem{b1}Asseo E., Pelletier G., Sol H., 1990, MNRAS, 247, 529
\bibitem{b2}Asseo E., Porzio A., 2006, MNRAS, 369, 1469
\bibitem{b3}Bartel N., Morris D., Sieber W., Hankins T. H., 1982, ApJ, 258, 776
\bibitem{b4}Bogdanov S., Grindlay J. E., 2009, ApJ, 703, 1557
\bibitem{b5}Burke-Spolaor S., Bailes M., 2010, MNRAS, 402, 855
\bibitem{b6}Buschauer R., Benford B., 1977, MNRAS, 179, 99
\bibitem{b7}Camilo F., Ransom S. M., Chatterjee S., Johnston S., Demorest P., 2012, %%@
ApJ, 746:63
\bibitem{b8}Deshpande A. A., Radhakrishnan V., 1992, J. Ap. Astr., 13, 151
\bibitem{b9}Gil J. A., Sendyk M., 2000, ApJ, 541, 351
\bibitem{b10}Gil J. A., Haberl F., Melikidze G., Geppert U., Zhang B., Melikidze G. %%@
Jr., 2008, ApJ, 686, 497
\bibitem{b11}Hankins T. H., Kern J. S., Weatherall J. C., Eilek J. A., 2003, Nat., 422, %%@
141
\bibitem{b12}Harding A. K., Muslimov A. G., 2001, ApJ, 556, 987
\bibitem{b13}Harding A. K., Muslimov A. G., 2002, ApJ, 568, 862
\bibitem{b14}Harding A. K., Muslimov A. G., 2011, ApJ, 743, 181
\bibitem{b15}Hibschman J. A., Arons J., 2001, ApJ, 554, 624
\bibitem{b16}Hoffman K., Heyl J., 2009, MNRAS, 400, 1986
\bibitem{b17}Ho W. C. G., Heinke C. O., 2009, Nat., 462, 71
\bibitem{b18}Jones P. B., 2010a, MNRAS, 401, 513 (Paper I)
\bibitem{b19}Jones P. B., 2010b, MNRAS, 409, 1719
\bibitem{b20}Jones P. B., 2011, MNRAS, 414, 759 (Paper II)
\bibitem{b21}Jones P. B., 2012, MNRAS, 419, 1682 (Paper III)
\bibitem{b22}Kaplan D. L., van Kerkwijk M. H., 2009, ApJ, 705, 798
\bibitem{b23}Keane E. F., McLaughlin M. A., 2011, Bull. Astr. Soc. India, 39, 333
\bibitem{b24}Keane E. F., Kramer M., Lyne A. G., Stappers B. W., McLaughlin M. A., %%@
2011, MNRAS, 415, 3065
\bibitem{b25}Kramer M., Lyne A. G., O'Brien J. T., Jordan C. A., Lorimer D. R., 2006, %%@
Sci., 312, 549
\bibitem{b26}Lyne A. G., 2009, in Becker W., ed., Astrophysics and Space Science %%@
Library, Vol. 357, Neutron Stars and Pulsars. Springer, Berlin, p.67
\bibitem{b27}Malofeev V. M., Gil J. A., Jessner A., Malov I. F., Seiradakis J. H., %%@
Sieber W., Wielebinski R., 1994, A\&A, 285, 201
\bibitem{b8}Manchester R. N., Hobbs G. B., Teoh A., Hobbs M., 2005, Astron J., 129, %%@
1993
\bibitem{b9}Medin Z., Lai D., 2006, Phys. Rev. A, 74, 062508
\bibitem{b30}Melrose D. B., 2000, in Kramer M., Wex N., Wielebinski N., eds, ASP Conf. 
Ser. Vol. 202, Pulsar Astronomy - 2000 and Beyond. Astron. Soc. Pac., San Francisco, %%@
p.721
\bibitem{b31}Michel F. C., 1974, ApJ, 192, 713
\bibitem{b32}Muslimov A. G., Tsygan A. I., 1992, MNRAS, 255, 61
\bibitem{b33}Palliyaguru N. T., et al, 2011, MNRAS, 417, 1871
\bibitem{b34}Pearson J. M., Goriely S., Chamel N., 2011, Phys. Rev. C, 83, 065810
\bibitem{b35}Rankin J. M., 1983, ApJ, 274, 333
\bibitem{b36}Rankin J. M., Wright G. A. E., 2008, MNRAS, 385, 1923
\bibitem{b37}Redman S. L., Rankin J. M., 2009, MNRAS, 395, 1529
\bibitem{b38}Ruderman M. A., Sutherland P. G., 1975, ApJ, 196, 51
\bibitem{b39}Soglasnov V. A., Popov M. V., Bartel N., Cannon W., Novikov A. Yu., %%@
Kondratiev V. I., Altunin V. I., 2004, ApJ, 616, 439
\bibitem{b40}Wang N., Manchester R. N., Johnston S., 2007, MNRAS, 377, 1383
\bibitem{b41}Weatherall J. C., 1997, ApJ, 483, 402
\bibitem{b42}Weatherall J. C., 1998, ApJ, 506, 341
\bibitem{b43}Weltevrede P., Edwards R. T., Stappers B. W., 2006, A\&A, 445, 243
\bibitem{b44}Weltevrede P., Stappers B. W., Rankin J. M., Wright G. A. E., 2006, ApJ, %%@
645, L149
\bibitem{b45}Yakovlev D. G., Ho W. C. G., Shternin P. S., Heinke C. O., Potekhin A. Y., %%@
2011, MNRAS, 411, 1977
\bibitem{b46}Yakovlev D. G., Pethick C. J., 2004, Ann. Rev. Astron. Ap., 42, 169
\bibitem{b47}Young M. D., Manchester R. N., Johnston S., 1999, Nat., 400, 848
\bibitem{b48}Zhu W., Kaspi V. M., Gonzalez M. E., Lyne A. G., 2009, ApJ, 704, 1321


\end{thebibliography}
\end{document}